\documentclass[aps,pre,reprint,groupedaddress]{revtex4-1}

\usepackage{graphicx}
\usepackage{amssymb}
\usepackage{amsmath}
\usepackage{docs}%
\usepackage{bm}%
\usepackage[colorlinks=true,linkcolor=blue]{hyperref}%
\usepackage{ subeqnarray}
\newcommand{\Peclet}{\textrm{Pe}}
\begin{document}


\title{Transient flows in active porous media}%

\author{Lefteris I. Kosmidis}
 \author{Kaare H. Jensen}%
\email{khjensen@fysik.dtu.dk}
\affiliation{Department of Physics, Technical University of Denmark, Kgs. Lyngby, DK-2800, Denmark}

\date{\today}

\begin{abstract}
Stimuli-responsive materials that modify their shape in response to changes in environmental conditions -- such as solute concentration, temperature, pH, and stress -- are widespread in nature and technology. Applications include micro- and nanoporous materials used in filtration and flow control. The physiochemical mechanisms that induce internal volume modifications have been widely studies. The coupling between induced volume changes and solute transport through porous materials, however, is not well understood. Here, we consider advective and diffusive transport through a small channel linking two large reservoirs. A section of stimulus-responsive material regulates the channel permeability, which is a function of the local solute concentration. We derive an exact solution to the coupled transport problem and demonstrate the existence of a flow regime in which the steady state is reached via a damped oscillation around the equilibrium concentration value. Finally, the feasibility of an experimental observation of the phenomena is discussed. Please note that this version of the paper has not been formally peer reviewed, revised or accepted by a journal. 
\end{abstract}

\maketitle


\section{Introduction}\label{intro} 
Fluid flow and convective solute transport in porous media and confined channel geometries are ubiquitous in nature and technology. Interesting phenomena arise when channels walls and solid structures are themselves active; for instance, when the presence of solutes influences the channel geometry and hence permeability to fluid flow. Man-made examples include sensing and actuation in microfluidic systems using stimuli-responsive hydrogels \cite{Koetting}. Responsive biomaterials are found, for example, in the phloem and xylem vascular systems of plants, where neighboring cells are separated by planar membranes covered with pores which respond to changes in concentration of chemical signals \cite{Zwieniecki2001,Mullendore2010}. The stimuli that induce changes in these synthetic and natural materials have been widely studies. However, the coupling between induced volume changes and advective solute transport in porous materials, however, are not well understood.

In this paper we investigate the transient nature of advective transport in active porous media. We study a one-dimensional system where the advective solute transport speed is coupled to the concentration field.  Numerical investigation of the model reveals the existence of a flow regime in which the steady state is reached via a damped oscillation around the equilibrium concentration value. We derive an exact solution using perturbation theory and show that the flow dynamics depends primarily on the ratio of advective to diffusive transport timescales (the Peclet number; $\Peclet$). Above a critical $\Peclet$-value, damped oscillations occur in both the velocity and concentration fields.
  Finally, we propose an experimental design to test the theoretical predictions.
\section{Flows in active porous media}
\begin{figure}
\includegraphics[width=8cm]{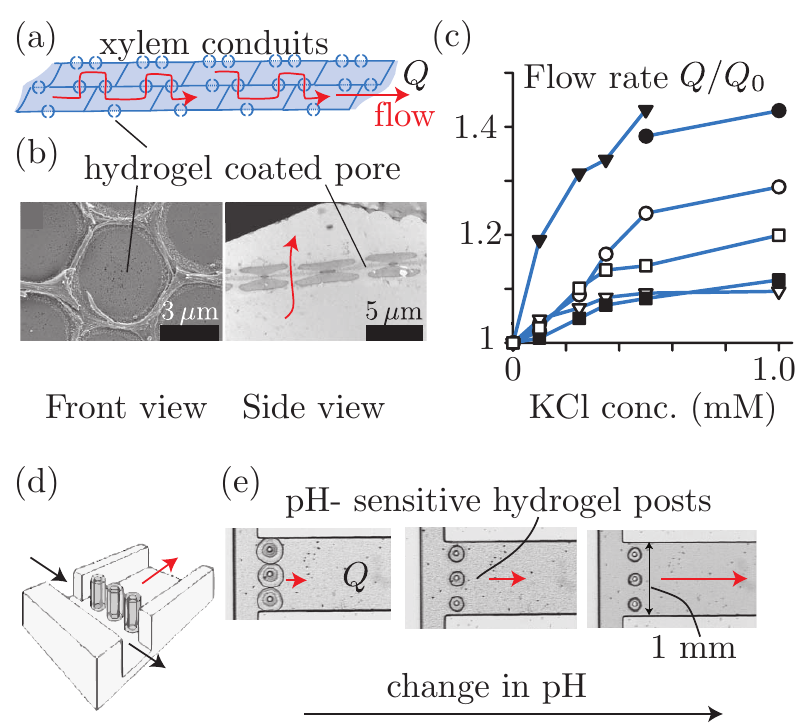}
\caption{Examples of active porous media in nature and technology. (a-b) Sketch and electron micrographs of hydrogel-coated pores  which separate xylem conduits in vascular plants. (c) The flow rate $Q$ through the conduits depends on the concentration of KCl, which influences the hydrogel permeability. Lines represent flow through different xylem samples.     (d-e) Sketch and micrographs of a concentration-dependent microfluidic valve. The valve comprises pH-sensitive hydrogel posts which shrink and swell in response to local pH. Panel (b) adapted from \cite{Choat2008}, (c) from \cite{Zwieniecki2001}, and (d-e) from  \citep{Beebe}. Reproduced with permission from copyright holders. \label{Fig:Fig1}}
\end{figure}

Stimulus-responsive hydrogels have been a topic of extensive research the past decades \cite{Koetting}. Their ability to modify their internal structure based on external stimuli allows for dynamic control over flows in biological  \cite{Zwieniecki2001,Mullendore2010} or man-made  systems \cite{Koetting,Peppas87} (Fig. \ref{Fig:Fig1}). Responsive hydrogels, i.e. hydrophilic polymers embedded and crosslinked into hydrophilic structures \cite{Peppas87}, can respond to a  broad range of stimuli, e.g. pH \cite{Gupta,Sharpe,Peppasph,Park,Chan1,Chan2,Kim,Qu,Schmaljohann}, temperature \cite{Klouda,Purushotham}, individual molecules (chemically driven) \cite{Wu,Bernfeld,Horbett,Klumb}, shear stress \cite{Hoffman,Qiu,Wang,Bell,Guvendiren,Gutowska} etc., that trigger a change of material properties. In pH induced responses, hydrogel swelling/deswelling occurs when polymers are ionized by the dynamically changing environmental pH \cite{Gupta}. Hence, the charge buildup results in an electrostatic force generation within the hydrogel that ultimately leads to absorbance or expulsion of water \cite{Sharpe,Peppasph}. Other workers have investigated temperature dependent hydrogels utilizing the critical solubility temperature with applications in drug delivery \cite{Klouda} and tissue engineering \cite{Purushotham}. Hydrogels also exhibit responsive behaviour to chemical stimuli such as glucose by entrapping glucose oxidase enzymes in the hydrogel structure \cite{Wu}. Another group of stimulus responsive hydrogels are known to respond to mechanical stress. Two subgroups that emerge are materials with shear thinning or shear thickening behaviour due to the viscoelastic nature of systems comprised of polymers, an intermediate material state at the interface between liquids and solids \cite{Hoffman,Qiu}. Applications of shear stress responsive hydrogels include, among others, drug delivery and wound repair \cite{Wang,Bell,Guvendiren,Gutowska}.

In summary, the physiochemical factors that induce volume chances in stimulus-responsive materials are well understood. By contrast, less is known about the coupling between fluid flow, solute advection and stimulus response in these systems. 

\section{Model}\label{model}
\begin{figure}
\includegraphics[width=8cm]{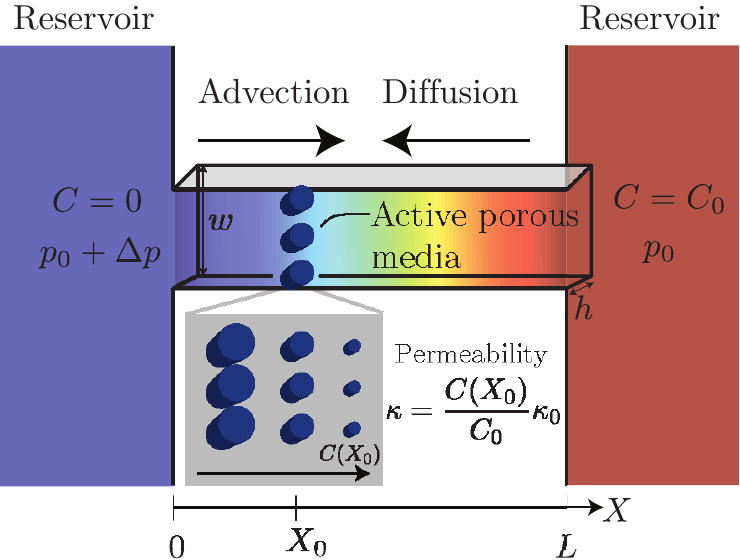}
\caption{Schematic illustration of the system. A small channel of length $L$, width $w$ and height $h$ link two large reservoirs. Differences in pressure ($\Delta p$) and solute concentration ($\Delta C= C_0$) drive diffusive and advective  transport of solute in opposite directions through the channel (large arrows). The active porous media pillars located at $X_0$ swell and shrink in inverse proportion to the local solute concentration (inset). This coupling between flow and concentration is modeled by concentration-dependent Darcy permeability $\kappa=\kappa_0 C(X_0)/C_0$.    \label{Fig:Fig2}}
\end{figure}

To elucidate the transient behavior of flow in active porous media, we consider flow in a small channel of constant cross section aligned with the horizontal $X$-axis linking two large reservoirs (Fig. \ref{Fig:Fig2}). The channel has length $L$, width $w$ and height $h$, and a short section of active porous media is located at $X=X_0<L$.
The right reservoir ($X=L$) is kept at constant concentration $X_0$, while the left reservoir ($X=0$) contains no solute. This drives a diffusive flux $-D\nabla C$ in the channel where $D$ is the diffusion coefficient. 
The right reservoir ($X=L$) is kept at constant pressure $p_0$, while the left reservoir ($x=0$) is at a higher pressure $p_0+\Delta p$. We assume the advective flow speed in the channel $v$ follows Darcy's law, $v\sim \kappa\Delta p/(\eta L)$, where $\kappa$ is the channel permeability and $\eta$ is the viscosity.
To model the active porous media, we assume that the dependence of the channel permeability $k$ on solute concentration $c$ can be expressed as $\kappa=\kappa_0 C(X_0)/C_0$. The hydraulic conductivity is proportional to the concentration at location the active porous media, $X_0$, such that high solute concentration evokes deswelling of the post valves while low concentration to an increase of post valve volume (Fig. \ref{Fig:Fig2}). 

The transport of solutes in the channel is governed by the advection-diffusion equation
\begin{equation}
\partial_T C + \mathbf v\cdot \nabla C = D\nabla^2C,
\end{equation}
where $T$ is time, $\mathbf v$ is the velocity field and $D$ is the diffusion coefficient. With the aforementioned assumptions, this reduces to a one-dimensional equation for the concentration $C(X,T)$ in the channel
\begin{equation}
\partial_T C + \frac{\kappa_0\Delta p}{\eta L}\frac{C(C_0)}{C_0}\partial_X C= D\partial_X^2C.
\end{equation}
The boundary conditions are
\begin{equation}
C(0)=0,\qquad C(L)=c_0. \label{eq:bc60}
\end{equation}
For convenience we introduce the non-dimensional variables 
\begin{equation}
x=X/L,\qquad c=C/C_0,\qquad \text{and}\qquad  t=T(D/L^2)
\end{equation}
The dimensionless governing equation is
 \begin{equation}
\partial_t c + \Peclet \, c(x_0) \partial_x c = \partial_x^2 c,\label{ge}
\end{equation}
where $x_0 = X_0/L$ and we have introduced the dimensionless Peclet number $\Peclet= v_0L/D $. Here, $v_0=\kappa_0\Delta p/(\eta L)$ is the maximum reference velocity. The Peclet number characterizes the relative contribution from advective and diffusive transport. The boundary conditions in Eq. \eqref{eq:bc60} become
\begin{equation}\label{bc}
c(0)=0, \qquad c(1)=1.
\end{equation}

\begin{figure*}\includegraphics[width=5.5cm]{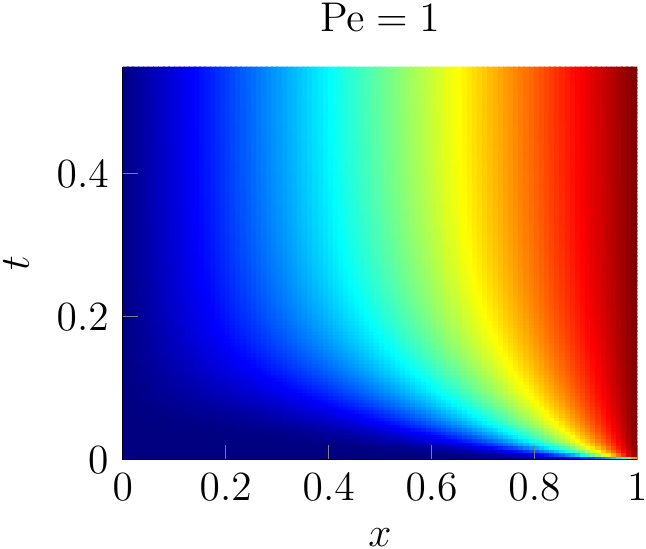}
\includegraphics[width=5.5cm]{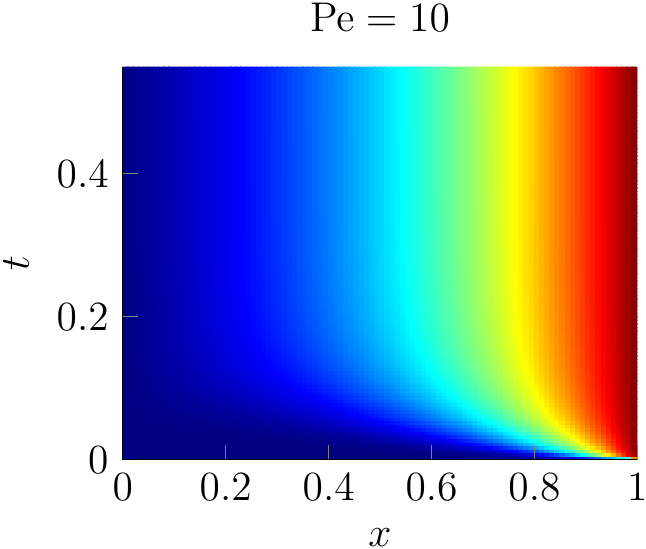}
\includegraphics[width=5.5cm]{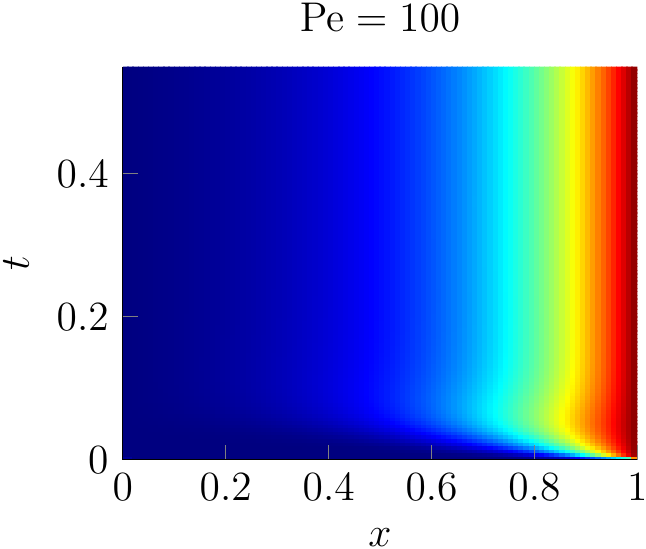}
	\caption{Oscillations in active porous media at high-$\Peclet$. Kymographs of the transient evolution of the solute concentration $c(x,t)$ in the pipe for Peclet numbers increasing from left to right $Pe=1, \, 10, \, 100$. The active porous media is located at $x_0  = 0.25$. Colors indicate concentration (dark blue: $c=0$, dark red: $c=1$). \label{Fig:Fig3}}
\end{figure*}

In the following, we consider the initial condition corresponding to an empty channel:
\begin{equation}
c(x,0)=0 \label{eq:initialEmpty}.
\end{equation}
and study the transient dynamics of the system. Before proceeding, however, we briefly discuss the steady-state solution $c_s(x)$ to Eq. \eqref{ge} and the system behavior when $\Peclet \ll 1$.
\subsection{Steady-state solution}
 When $\partial_tc =0$, Eq. \eqref{ge} reduces to
\begin{equation}\label{ss}
\Peclet \, \gamma \partial_x c_s(x) = \partial_x^2 c_s (x), 
\end{equation}
where we have introduced the parameter $\gamma=c_s(x_0)$, the steady-state concentration at $x_0$. The solution to Eq. \eqref{ss} with boundary conditions \eqref{bc} is
\begin{equation}\label{sssolution}
c_s(x) = \frac{e^{\Peclet \, \gamma x}-1}{e^{\Peclet \, \gamma }-1}.
\end{equation}
The parameter $\gamma=c_s(x_0)$ can be determined as a function of the system parameters ($Pe$ and $x_0$) by solving the trancendental equation
\begin{equation}\label{gamma}
\gamma = \frac{e^{\Peclet \, \gamma x_0}-1}{e^{\Peclet \, \gamma }-1}.
\end{equation}
When $\Peclet\ll 1$, we find that $c(x) = x$ and $\gamma = x_0$.  Taking the limit $\Peclet \gg 1$ leads to $\log \gamma /\gamma =\Peclet(x_0-1)$ 
\subsection{Solution in a diffusion dominated system}
When diffusion dominates, the Peclet number $\Peclet \ll 1$. In that limit, the solution to Eqns. \eqref{ge}-\eqref{bc} is
\begin{equation}
c(x,t) = x+\frac 2\pi \sum_n\frac{(-1)^n}{n}\exp^{-n^2\pi^2 t}\sin (n\pi x) \label{eq:diffsol}.
\end{equation}
Equilibrium is approached exponentially on the timescale set by the slowest mode ($n=1$). The system is within $5\%$ of the steady state solution when $t \simeq  3 \pi^{-2}\simeq 0.3$.

\section{Results}
\subsection{Numerical Simulation}
In order to reveal the transient nature of flow in active porous media (Fig. \ref{Fig:Fig2}), we ran simulations of Eqns. \eqref{ge}-\eqref{eq:initialEmpty} for a range of values for $\textit{Pe}$ and $x_0$.
For relatively low Peclet numbers -- corresponding to a diffusion-dominated system -- the steady state is reached asymptotically with non-dimensional relaxation time $\tau\sim 0.3$. The behavior of the system is thus in accord with a purely diffusive process (Eq. \eqref{eq:diffsol}), where equilibrium is approached exponentially on a similar timescale.

By contrast, for values of the Peclet number $\Peclet$ above unity, the characteristics of the system changes in two respects. First, the steady state is reached on a timescale which decreases with increasing $\Peclet$. Second, for large $\Peclet$ the approach to equilibrium follows a damped oscillation (Fig. \ref{Fig:Fig3}(c)), indicating a qualitative deviation from the asymptotic approach to equilibrium in Eq. \eqref{eq:diffsol} and Fig. \ref{Fig:Fig3}(a). 

To further elucidate the characteristics of the oscillations, we studied the temporal evolution of a small disturbance to the steady state. We thus added a weak gaussian perturbation to the steady-state solution (Eq. \eqref{sssolution}) at an arbitrary position $x_p$ within the domain, and studied the approach to equilibrium. After the initial  perturbation had decayed, we observed an approximately decaying harmonic time dependence of the disturbance at the position $x_{\text{obs}}$, i.e. 
\begin{equation}
c(x_{\text{obs}},t) - c_s(x_{\text{obs}}) \propto e^{-(k_r + i k_i)t}\label{eq:fit}
\end{equation}
where $c_s$ is the steady state concentration given in Eq.~\eqref{sssolution}.
In Eq. \eqref{eq:fit}, $k_r$ and $k_i$ are the real and imaginary parts of the complex wavenumber $k$, corresponding to decay time $\sim  k_r^{-1}$ and oscillation period $\sim 2\pi k_i^{-1}$.  We thus extracted $k_r$ and $k_i$ from the numerical simulations by curve fitting using.
Neither the position of the perturbation $x_p$ nor the observation location $x_{\text{obs}}$ appeared to influence the magnitude of the wavenumber $k$ significantly. However, we chose the parameters to avoid overlap between the position of the active porous material $x_0$ and $x_p$ and $x_{\text{obs}}$. Finally, we found that while oscillations are present when the position is to the right of the channel centerline ($x_0>1/2$), they decay rapidly and a sum of at least two decaying exponentials are necessary to provide a satisfactory curve fit. In the following, we thus restrict ourselves to the case $x_0<1/2$.

\begin{figure}[t!]
		\includegraphics[width=8cm]{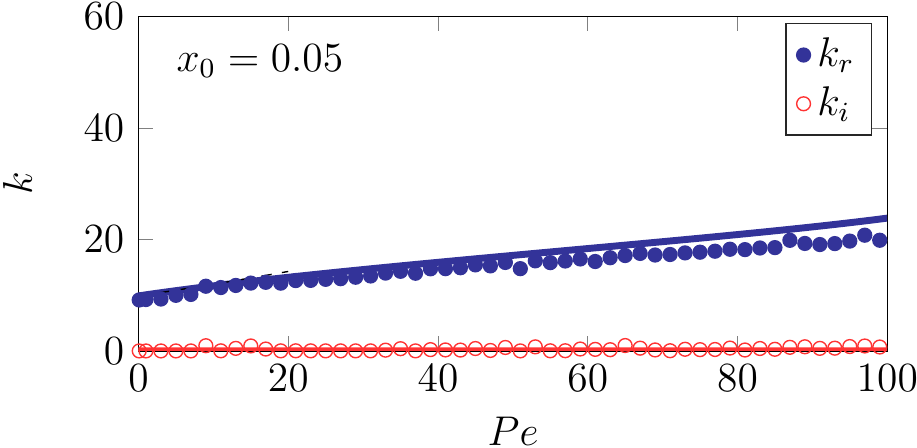}
				\includegraphics[width=8cm]{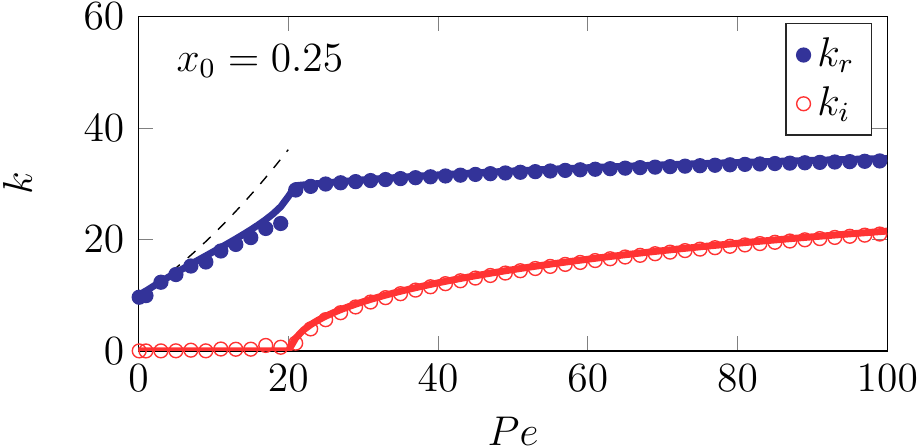}
						\includegraphics[width=8cm]{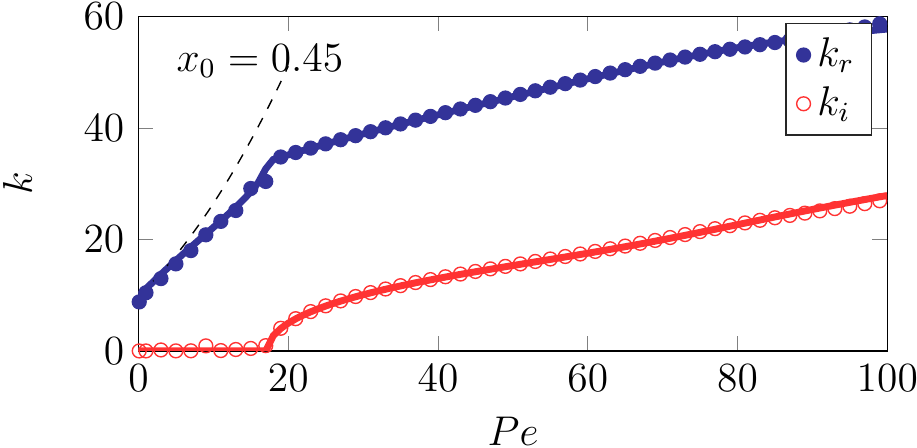}
\caption{Onset of oscillations in active porous media with increasing $\Peclet$. The real (dots) and imaginary (circle) wavenumber $k$ plotted as a function of Peclet number $\textit{Pe}$ for $x_0=0.05,\;0.25,\; 0.45$. For $x_0=0.25$ the oscillations appear at $\Peclet\simeq 20$ where the first non-zero $k_i$ is found. Data points were determined from fits to numerical data (Fig. \ref{Fig:Fig3}) using Eq. \eqref{eq:fit}. Thick solid lines show results from the analytical solution in Eq. \eqref{eq:eig}. Thin dashed lines indicate the low-$\Peclet$ limit in Eq. \eqref{eq:ksmall}.
 \label{Fig:Fig4}}
\end{figure}	

Having extracted the wavenumber $k=k_r+ik_i$ from the numerical simulations, we studied $k$'s dependence on the relative importance of advection and diffusion (Fig. \ref{Fig:Fig4}). When the Peclet number is relatively small, we found $k_r\simeq 10$ and $k_i=0$, in accord with Eq. \eqref{eq:diffsol}, which predicts $k_r = 1^2 \pi ^2\simeq 9.87$ and $k_i=0$.
The simulations further revealed that the onset of oscillations occurs at a critical value of the Peclet number $\Peclet_c$. For the case $x_0=1/4$ shown in Fig. \ref{Fig:Fig4}, $\Peclet_c \simeq  20$. Note that the magnitude of the critical $\Peclet_c$ varies depending on the location $x_0$ of the active porous media in the channel (see also Fig. \ref{Fig:Fig6}).

The physical mechanism that triggers the onset of oscillations can be interpreted as follows: when the system is perturbed away from the steady state $c_s(x)$, the concentration at $x_0$, i.e. $c(x_0)$, will shift either up or down as solute is transported by a convention through the domain. This directly influences the advective flow speed, which is proportional to the local concentration at that point, $v\propto c(x_0)/c_0$. Diffusion will counteract this process, eventually returning the system to the steady state $c_s(x)$.
However, if the advective transport is sufficiently strong, advection can push the system into a state in where the concentration gradients become so great that the concentration $c(x_0)$ overshoots it's equilibrium value as diffusion counteracts advection. The process repeats itself -- with a progressively smaller amplitude -- until the steady state is restored.

\subsection{Analytic Solution}

To rationalize the observed onset of oscillations at high Peclet-numbers (Fig. \ref{Fig:Fig3} and \ref{Fig:Fig4}) and their dependence on the system parameters, we proceed to consider the evolution of the perturbed system. Considering a small deviation from the steady state $c_s(x)$, we write 
\begin{equation}
c(x,t) = c_s(x) - c_1(x,t),\label{eq:perturb}
\end{equation}
 where we assume the perturbation $c_1 \ll c_s$. We further assume that the perturbation has a harmonic time dependence
\begin{equation}
c_1(x,t) = e^{-kt} g(x), \label{eq:perturbForm}
\end{equation}
where $k = k_{r} + ik_{i}$ is the complex wavenumber and $k_{r}>0$ and $g(x)$ is an unknown function of $x$. Substitution of Eq. \eqref{eq:perturbForm}-\eqref{eq:perturb} into Eq. (\ref{ge}) leads to a spatial equation for $g(x)$\begin{equation}
\label{g}
g''(x) - \Peclet \, \gamma g'(x) +kg(x) = \frac{ \Peclet^2 \, \gamma}{e^{\Peclet \, \gamma }-1}e^{\Peclet \, \gamma x} g(x_0),
\end{equation}
where prime denotes derivative with respect to $x$. The boundary conditions are
\begin{equation}
g(0)=0, \quad g(1)=0, \quad g(x_0)=1
\end{equation}
where we have eliminated quadratic terms in $g$. Note that $g(x_0)$ is an arbitrary constant that defines the strength of the perturbation, chosen here as unity. 

Equation (\ref{g}) is solved following the method of \citet{Pedley1978}, who analyzed a similar problem related to transient flows near osmotic membranes. A particular solution to the inhomogeneous equation is $g_{i}(x) = \lambda \exp\left(\Peclet^* x\right)$, while the homogeneous solution is $g_{h}=\exp\left(\Peclet^*x/2\right)(A\cos \zeta x + B \sin \zeta x)$.
Here, we have introduced the parameters 
\begin{subeqnarray}
\Peclet^* &=& \Peclet \, \gamma,\\
\lambda &=& \frac 1k  \frac{\Peclet^{*2}}{\gamma(e^{\Peclet^*}-1)},\\
\zeta &=& \frac{1}{2}\Peclet^* \sqrt{\frac{4}{\lambda \gamma(e^{\Peclet^*}-1)}-1}.
\end{subeqnarray}
The complete solution to \eqref{g} is
\begin{equation}
g(x) = \lambda e^{\Peclet^* x}+e^{\frac{\Peclet^* \, x}{2}}(A\cos \zeta x + B \sin \zeta x)
\end{equation}
To determine the constants $A$ and $B$ we apply the boundary conditions
\begin{equation}
\label{bc1}
g(0) = 0, \quad g(1) = 0, \quad g(x_0) = 1,
\end{equation}
which after substitution become
\begin{subeqnarray}
\label{bc2}
\lambda+ A &= &0 \\
\lambda e^{\Peclet^*} + e^{\frac{\Peclet^*}{2}}(A \cos \zeta + B \sin \zeta) &=& 0 \\
\lambda e^{\Peclet^* \, x_0} + e^{\frac{\Peclet^* \, x_0}{2}}(A\cos \zeta x_0 + B \sin \zeta x_0) &=& 1\qquad
\end{subeqnarray}
By eliminating $A$ and $B$ we find an eigenvalue equation for the wavenumber $k$:
\begin{equation}
\lambda e^{\frac{\Peclet^* \, x_0}{2}} \left[e^{\frac{\Peclet^* \, x_0}{2}} - \cos \zeta x_0 +  \frac{\sin \zeta x_0}{\sin{\zeta}}\left(\cos \zeta-e^{\frac{\Peclet^*}{2}}\right) \right] = 1.\label{eq:eig}
\end{equation}

\begin{figure}
\includegraphics[width=8cm]{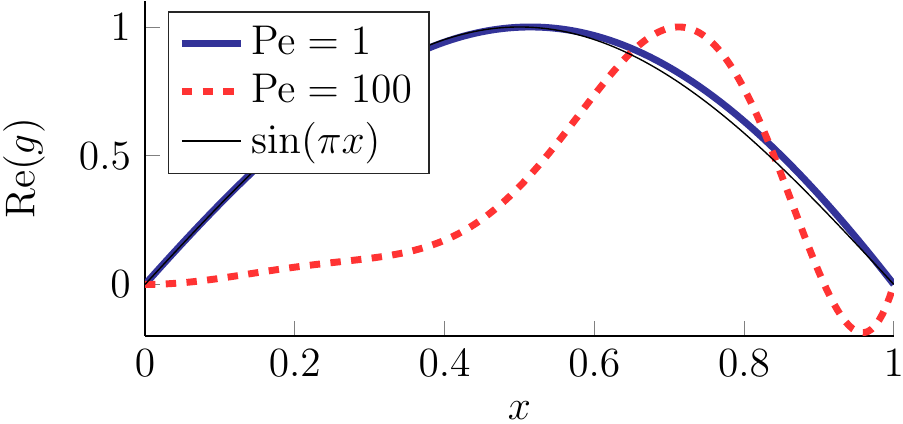}\\
\includegraphics[width=8cm]{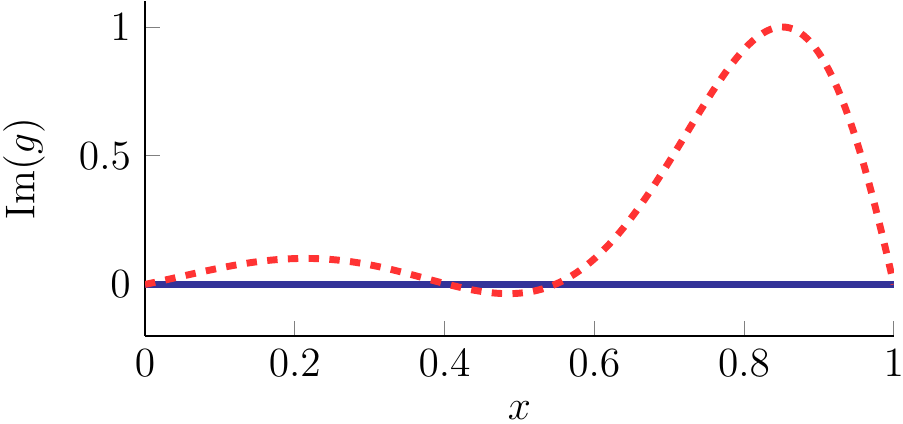}
 \caption{Real and imaginary parts of the eigenfunctions corresponding to Eq. \eqref{funct} for $x_0=0.25$ shown for $\Peclet = 1$ (thick solid line) and $\Peclet=100$ (dashed line). The real eigenfunction  approaches $g\sim \sin \pi x$ in limit $\Peclet \to 0$ (thin solid line), in accord with Eq. \eqref{eq:diffsol}\label{Fig:Fig5}. }
\end{figure}

To test the validity of our solution, we compared the predictions from Eq. \eqref{eq:eig} with numerical data. 
For a given set of parameters $(\Peclet, x_0)$ we thus determined the solution to Eq. \eqref{eq:eig} with the smallest real part of $k$, corresponding to the slowest decaying mode. The solutions to Eq. \eqref{eq:eig} are in good agreement with the numerically extracted eigenvalues (Fig. \ref{Fig:Fig4}).
The eigenfunction is 
\begin{multline}
\label{funct}
g(x) = \lambda \bigg[ e^{\Peclet^* \, x}
+ e^{\frac{\Peclet^* \, x}{2}}\bigg(-\cos \zeta x+ \\
+ \bigg(\cot \zeta - \frac{e^{\frac{\Peclet^*}{2}}}{\sin \zeta}\bigg)\sin \zeta x \bigg) \bigg],
\end{multline}
shown in Fig. ~\ref{Fig:Fig5}. We note that the spatial eigenfunctions in Eq. \eqref{funct} are consistent with Eq. \eqref{eq:diffsol} when $\Peclet$ is relatively small.

\subsection{Critical \Peclet{}  for onset of oscillations}
\begin{figure}[]
		\includegraphics[width=8cm]{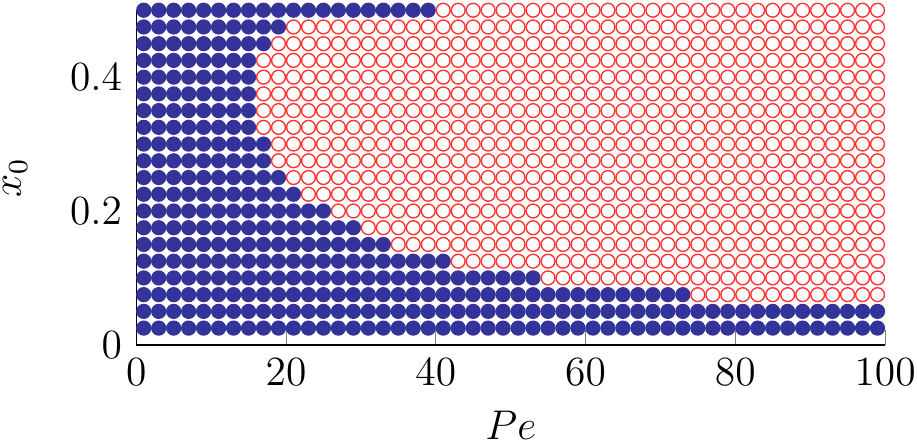}
				\includegraphics[width=8cm]{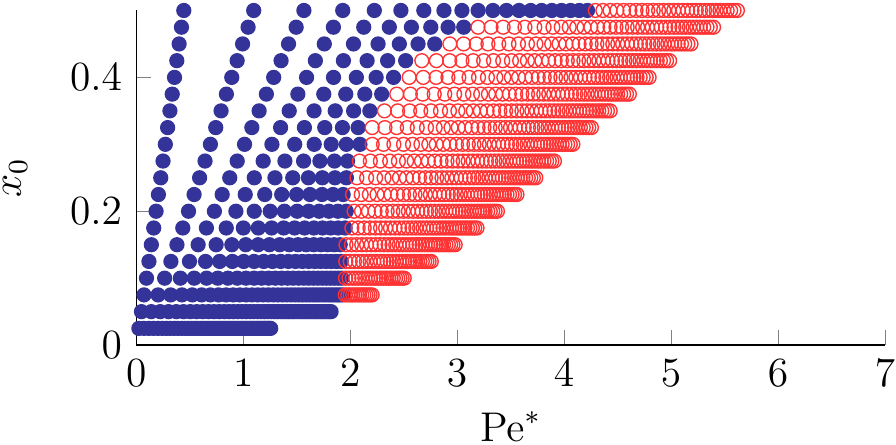}
	\caption{Phase diagrams showing the onset of oscillations at or above $\Peclet \simeq 18$ and $\Peclet^*\simeq 2$ depending on the location of the active porous material $x_0$. Blue dots indicate the absence of oscillations ($k_i=0$) while red circles corresponds to parameter choices at which oscillations are observed. \label{Fig:Fig6}}
\end{figure}

To elucidate the conditions under which damped oscillations occur in our system,
we extracted a phase diagram (Fig. \ref{Fig:Fig6}) from the eigenvalue equation \eqref{eq:eig}. Oscillations in the mode associated with the smallest real eigenvalue can occur for values for the Peclet number at or above $\Peclet \simeq 18$, depending on the position of the active porous media $x_0$. This suggests that advection should be nearly twenty times stronger than diffusion to obtain oscillations. However, because of the coupling between the permeability of the porous media $\kappa= \kappa_0 c(x_0)/c_0=\kappa \gamma$ and concentration $c(x_0)$, we can write for the flow speed $v = \kappa \Delta p/(\eta L)=\gamma v_0$. This implies that the physically relevant Peclet number is $\Peclet^*$, given by
\begin{equation}
\Peclet^* = \frac{v}{v_0}\Peclet= \gamma \Peclet
\end{equation}
Replotting the phase diagram using the rescaled Peclet number $\Peclet^*$ reveals that the onset of oscillations occur when advection is $2$ to $4$ times stronger than diffusion.

Oscillations are found in the numerical simulations for $x_0>1/2$. As noted earlier, however, they decay rapidly and a sum of at least two decaying exponentials are necessary to provide a satisfactory curve fit. By extracting the three smallest roots of Eq. \eqref{eq:eig}, we found that for $x_0>1/2$ the oscillations are no longer associated with mode with the smallest real eigenvalue. Oscillations are found, however, in higher-order solutions to Eq. \eqref{eq:eig}, an observation which provides a qualitative rational for the numerical results.

\subsection{Small-\Peclet{}  expansion}
We end this section by deriving an analytical expression for the solution to Eq. \eqref{eq:eig} for small $\Peclet$. Taking the limit $\Peclet \ll 1$ in Eq. \eqref{bc2} leads to $\lambda =\Peclet/k$ and $\zeta=\sqrt{k}$. Interting this into Eq. \eqref{eq:eig} and assuming that we can write the eigenvalue as  a power-series in $\Peclet$: $k^{1/2} = a_0 + \Peclet\, a_1 + \mathcal O(\Peclet^2)$ with $a_0=\pi$ gives an analytical expression for the eigenvalue $k$ at low-$\Peclet$
\begin{equation}
k(Pe,x_0)^{1/2} = \pi + 2\Peclet\frac{\sin(\pi x_0)}{\pi^2}+\mathcal O(\Peclet^2).\label{eq:ksmall}
\end{equation}
The approximate expression in Eq. \eqref{eq:ksmall} is in reasonable accord with the solution to Eq. \eqref{eq:eig} for $Pe\lesssim 10$ (Fig. \ref{Fig:Fig4}).

\section{Discussion and conclusion}
A relatively complete picture of the factors that influence transient flows in active porous media has emerged. First and foremost, we have demonstrated the existence of damped oscillations in the flow within a channel linking two reservoirs (Fig. \ref{Fig:Fig2}). The oscillations in solute concentration and liquid flow speed is the result of a coupling between solute concentration $c$ and the permeability of the channel, controlled by the swelling and shrinking of a stimulus-responsive material located at the position $x_0$ in the channel. Damped oscillations occur when the advective transport $\sim v c$ is sufficiently great to overcome diffusive transport $\sim D c/L$, i.e. when the Peclet number $\Peclet^* = vL/D$ is greater than a critical value, which varies in the range from $\Peclet^*\simeq 2 $ to $\Peclet^*\simeq 4$, depending on the location of the active porous material $x_0$ (Fig. \ref{Fig:Fig6}).
 
 To observe the damped oscillations in a laboratory setting, we an experiment based on the system in Fig. \ref{Fig:Fig2}. In a channel of length $L=1$ mm, the characteristic diffusive time is $t_d=L^2/D= 2\times 10^3$ s, where we have used the diffusion coefficient $D=5\times 10^{-10}$ m$^2$/s of the dye carboxyflourescein \cite{Carroll2014}. With a stimulus-responsive hydrogel valve located at $x_0=0.25$ mm operated at a moderate Peclet number of $\Peclet^* = 4$ ($\Peclet = 100$), we find $k_r\sim 30$ and $k_i\sim 20$. This corresponds to an oscillation period of $T_{\text{osc}}=2\pi k_i^{-1}t_d\simeq 630$ s while the decay time is $\tau = 3 t_d k_r^{-1}\simeq 200$ s. At least the first half-period should be observable for this choice of parameters. For a slightly lower forcing $\Peclet^* = 2$ ($\Peclet\simeq 30$), we find $k_r\sim 30$ and $k_i\sim 10$ corresponding to oscillation period $T_{\text{osc}}\simeq 1300$ s and decay time $\tau \simeq 200$ s.
For water flowing in a channel of width $w = 300$ $\mathrm{\mu m}$ and height $h = 100$ $\mathrm{\mu m}$, these flows would require pressure differentials of $\Delta p = 0.077$ Pa and $\Delta p = 0.023$ Pa respectively \cite{bruus}. These pressure differentials, in the absence of post valves would generate flow velocities $v_0=\Peclet D/L = 50$ $\mathrm{\mu m /s}$ and $v_0 = 15$ $\mathrm{\mu m /s}$ for $\Peclet = 100$ and $\Peclet = 30$ respectively, however due to the reduced conductance imposed by the hydrogel structures, the effective velocity for $\Peclet^*=4$ is $v = 2$ $\mathrm{\mu m /s}$ and for $\Peclet^*=2$, $v = 1$ $\mathrm{\mu m /s}$. In summary, it does not appear technically unfeasible to experimentally validate the existence of damped oscillations in active porous media.


\begin{thebibliography}{30}%
\makeatletter
\providecommand \@ifxundefined [1]{%
 \@ifx{#1\undefined}
}%
\providecommand \@ifnum [1]{%
 \ifnum #1\expandafter \@firstoftwo
 \else \expandafter \@secondoftwo
 \fi
}%
\providecommand \@ifx [1]{%
 \ifx #1\expandafter \@firstoftwo
 \else \expandafter \@secondoftwo
 \fi
}%
\providecommand \natexlab [1]{#1}%
\providecommand \enquote  [1]{``#1''}%
\providecommand \bibnamefont  [1]{#1}%
\providecommand \bibfnamefont [1]{#1}%
\providecommand \citenamefont [1]{#1}%
\providecommand \href@noop [0]{\@secondoftwo}%
\providecommand \href [0]{\begingroup \@sanitize@url \@href}%
\providecommand \@href[1]{\@@startlink{#1}\@@href}%
\providecommand \@@href[1]{\endgroup#1\@@endlink}%
\providecommand \@sanitize@url [0]{\catcode `\\12\catcode `\$12\catcode
  `\&12\catcode `\#12\catcode `\^12\catcode `\_12\catcode `\%12\relax}%
\providecommand \@@startlink[1]{}%
\providecommand \@@endlink[0]{}%
\providecommand \url  [0]{\begingroup\@sanitize@url \@url }%
\providecommand \@url [1]{\endgroup\@href {#1}{\urlprefix }}%
\providecommand \urlprefix  [0]{URL }%
\providecommand \Eprint [0]{\href }%
\providecommand \doibase [0]{http://dx.doi.org/}%
\providecommand \selectlanguage [0]{\@gobble}%
\providecommand \bibinfo  [0]{\@secondoftwo}%
\providecommand \bibfield  [0]{\@secondoftwo}%
\providecommand \translation [1]{[#1]}%
\providecommand \BibitemOpen [0]{}%
\providecommand \bibitemStop [0]{}%
\providecommand \bibitemNoStop [0]{.\EOS\space}%
\providecommand \EOS [0]{\spacefactor3000\relax}%
\providecommand \BibitemShut  [1]{\csname bibitem#1\endcsname}%
\let\auto@bib@innerbib\@empty
\bibitem [{\citenamefont {Koetting}\ \emph {et~al.}(2015)\citenamefont
  {Koetting}, \citenamefont {Peters}, \citenamefont {Steichen},\ and\
  \citenamefont {Peppas}}]{Koetting}%
  \BibitemOpen
  \bibfield  {author} {\bibinfo {author} {\bibfnamefont {M.~C.}\ \bibnamefont
  {Koetting}}, \bibinfo {author} {\bibfnamefont {J.~T.}\ \bibnamefont
  {Peters}}, \bibinfo {author} {\bibfnamefont {S.~D.}\ \bibnamefont
  {Steichen}}, \ and\ \bibinfo {author} {\bibfnamefont {N.~A.}\ \bibnamefont
  {Peppas}},\ }\href@noop {} {\bibfield  {journal} {\bibinfo  {journal}
  {Materials Science and Engineering: R: Reports}\ }\textbf {\bibinfo {volume}
  {93}},\ \bibinfo {pages} {1 } (\bibinfo {year} {2015})}\BibitemShut {NoStop}%
\bibitem [{\citenamefont {Zwieniecki}\ \emph {et~al.}(2001)\citenamefont
  {Zwieniecki}, \citenamefont {Melcher},\ and\ \citenamefont
  {Holbrook}}]{Zwieniecki2001}%
  \BibitemOpen
  \bibfield  {author} {\bibinfo {author} {\bibfnamefont {M.~A.}\ \bibnamefont
  {Zwieniecki}}, \bibinfo {author} {\bibfnamefont {P.~J.}\ \bibnamefont
  {Melcher}}, \ and\ \bibinfo {author} {\bibfnamefont {N.~M.}\ \bibnamefont
  {Holbrook}},\ }\href@noop {} {\bibfield  {journal} {\bibinfo  {journal}
  {science}\ }\textbf {\bibinfo {volume} {291}},\ \bibinfo {pages} {1059}
  (\bibinfo {year} {2001})}\BibitemShut {NoStop}%
\bibitem [{\citenamefont {Mullendore}\ \emph {et~al.}(2010)\citenamefont
  {Mullendore}, \citenamefont {Windt}, \citenamefont {Van~As},\ and\
  \citenamefont {Knoblauch}}]{Mullendore2010}%
  \BibitemOpen
  \bibfield  {author} {\bibinfo {author} {\bibfnamefont {D.~L.}\ \bibnamefont
  {Mullendore}}, \bibinfo {author} {\bibfnamefont {C.~W.}\ \bibnamefont
  {Windt}}, \bibinfo {author} {\bibfnamefont {H.}~\bibnamefont {Van~As}}, \
  and\ \bibinfo {author} {\bibfnamefont {M.}~\bibnamefont {Knoblauch}},\
  }\href@noop {} {\bibfield  {journal} {\bibinfo  {journal} {The Plant Cell}\
  }\textbf {\bibinfo {volume} {22}},\ \bibinfo {pages} {579} (\bibinfo {year}
  {2010})}\BibitemShut {NoStop}%
\bibitem [{\citenamefont {Huglin}(1989)}]{Peppas87}%
  \BibitemOpen
  \bibfield  {author} {\bibinfo {author} {\bibfnamefont {M.~R.}\ \bibnamefont
  {Huglin}},\ }\href@noop {} {\bibfield  {journal} {\bibinfo  {journal}
  {British Polymer Journal}\ }\textbf {\bibinfo {volume} {21}},\ \bibinfo
  {pages} {184} (\bibinfo {year} {1989})}\BibitemShut {NoStop}%
\bibitem [{\citenamefont {Gupta}\ \emph {et~al.}(2002)\citenamefont {Gupta},
  \citenamefont {Vermani},\ and\ \citenamefont {Garg}}]{Gupta}%
  \BibitemOpen
  \bibfield  {author} {\bibinfo {author} {\bibfnamefont {P.}~\bibnamefont
  {Gupta}}, \bibinfo {author} {\bibfnamefont {K.}~\bibnamefont {Vermani}}, \
  and\ \bibinfo {author} {\bibfnamefont {S.}~\bibnamefont {Garg}},\ }\href@noop
  {} {\bibfield  {journal} {\bibinfo  {journal} {Drug Discovery Today}\
  }\textbf {\bibinfo {volume} {7}},\ \bibinfo {pages} {569 } (\bibinfo {year}
  {2002})}\BibitemShut {NoStop}%
\bibitem [{\citenamefont {Sharpe}\ \emph {et~al.}(2014)\citenamefont {Sharpe},
  \citenamefont {Daily}, \citenamefont {Horava},\ and\ \citenamefont
  {Peppas}}]{Sharpe}%
  \BibitemOpen
  \bibfield  {author} {\bibinfo {author} {\bibfnamefont {L.~A.}\ \bibnamefont
  {Sharpe}}, \bibinfo {author} {\bibfnamefont {A.~M.}\ \bibnamefont {Daily}},
  \bibinfo {author} {\bibfnamefont {S.~D.}\ \bibnamefont {Horava}}, \ and\
  \bibinfo {author} {\bibfnamefont {N.~A.}\ \bibnamefont {Peppas}},\
  }\href@noop {} {\bibfield  {journal} {\bibinfo  {journal} {Expert Opinion on
  Drug Delivery}\ }\textbf {\bibinfo {volume} {11}},\ \bibinfo {pages} {901}
  (\bibinfo {year} {2014})}\BibitemShut {NoStop}%
\bibitem [{Pep(2000)}]{Peppasph}%
  \BibitemOpen
  \href@noop {} {\bibfield  {journal} {\bibinfo  {journal} {European Journal of
  Pharmaceutics and Biopharmaceutics}\ }\textbf {\bibinfo {volume} {50}},\
  \bibinfo {pages} {27 } (\bibinfo {year} {2000})}\BibitemShut {NoStop}%
\bibitem [{\citenamefont {Park}\ \emph {et~al.}(1998)\citenamefont {Park},
  \citenamefont {Song}, \citenamefont {Kim},\ and\ \citenamefont {Kim}}]{Park}%
  \BibitemOpen
  \bibfield  {author} {\bibinfo {author} {\bibfnamefont {H.-Y.}\ \bibnamefont
  {Park}}, \bibinfo {author} {\bibfnamefont {I.-H.}\ \bibnamefont {Song}},
  \bibinfo {author} {\bibfnamefont {J.-H.}\ \bibnamefont {Kim}}, \ and\
  \bibinfo {author} {\bibfnamefont {W.-S.}\ \bibnamefont {Kim}},\ }\href@noop
  {} {\bibfield  {journal} {\bibinfo  {journal} {International Journal of
  Pharmaceutics}\ }\textbf {\bibinfo {volume} {175}},\ \bibinfo {pages} {231 }
  (\bibinfo {year} {1998})}\BibitemShut {NoStop}%
\bibitem [{\citenamefont {Chan}\ \emph {et~al.}(2008)\citenamefont {Chan},
  \citenamefont {Whitney},\ and\ \citenamefont {Neufeld}}]{Chan1}%
  \BibitemOpen
  \bibfield  {author} {\bibinfo {author} {\bibfnamefont {A.~W.}\ \bibnamefont
  {Chan}}, \bibinfo {author} {\bibfnamefont {R.~A.}\ \bibnamefont {Whitney}}, \
  and\ \bibinfo {author} {\bibfnamefont {R.~J.}\ \bibnamefont {Neufeld}},\
  }\href@noop {} {\bibfield  {journal} {\bibinfo  {journal}
  {Biomacromolecules}\ }\textbf {\bibinfo {volume} {9}},\ \bibinfo {pages}
  {2536} (\bibinfo {year} {2008})}\BibitemShut {NoStop}%
\bibitem [{\citenamefont {Chan}\ and\ \citenamefont {Neufeld}(2009)}]{Chan2}%
  \BibitemOpen
  \bibfield  {author} {\bibinfo {author} {\bibfnamefont {A.~W.}\ \bibnamefont
  {Chan}}\ and\ \bibinfo {author} {\bibfnamefont {R.~J.}\ \bibnamefont
  {Neufeld}},\ }\href@noop {} {\bibfield  {journal} {\bibinfo  {journal}
  {Biomaterials}\ }\textbf {\bibinfo {volume} {30}},\ \bibinfo {pages} {6119 }
  (\bibinfo {year} {2009})}\BibitemShut {NoStop}%
\bibitem [{\citenamefont {Kim}\ \emph {et~al.}(2000)\citenamefont {Kim},
  \citenamefont {Cho}, \citenamefont {Lee},\ and\ \citenamefont {Kim}}]{Kim}%
  \BibitemOpen
  \bibfield  {author} {\bibinfo {author} {\bibfnamefont {S.~Y.}\ \bibnamefont
  {Kim}}, \bibinfo {author} {\bibfnamefont {S.~M.}\ \bibnamefont {Cho}},
  \bibinfo {author} {\bibfnamefont {Y.~M.}\ \bibnamefont {Lee}}, \ and\
  \bibinfo {author} {\bibfnamefont {S.~J.}\ \bibnamefont {Kim}},\ }\href@noop
  {} {\bibfield  {journal} {\bibinfo  {journal} {Journal of Applied Polymer
  Science}\ }\textbf {\bibinfo {volume} {78}},\ \bibinfo {pages} {1381}
  (\bibinfo {year} {2000})}\BibitemShut {NoStop}%
\bibitem [{\citenamefont {Qu}\ \emph {et~al.}(2000)\citenamefont {Qu},
  \citenamefont {Wirsan},\ and\ \citenamefont {Albertsson}}]{Qu}%
  \BibitemOpen
  \bibfield  {author} {\bibinfo {author} {\bibfnamefont {X.}~\bibnamefont
  {Qu}}, \bibinfo {author} {\bibfnamefont {A.}~\bibnamefont {Wirsan}}, \ and\
  \bibinfo {author} {\bibfnamefont {A.-C.}\ \bibnamefont {Albertsson}},\
  }\href@noop {} {\bibfield  {journal} {\bibinfo  {journal} {Polymer}\ }\textbf
  {\bibinfo {volume} {41}},\ \bibinfo {pages} {4589 } (\bibinfo {year}
  {2000})}\BibitemShut {NoStop}%
\bibitem [{\citenamefont {Schmaljohann}(2006)}]{Schmaljohann}%
  \BibitemOpen
  \bibfield  {author} {\bibinfo {author} {\bibfnamefont {D.}~\bibnamefont
  {Schmaljohann}},\ }\href@noop {} {\bibfield  {journal} {\bibinfo  {journal}
  {Advanced Drug Delivery Reviews}\ }\textbf {\bibinfo {volume} {58}},\
  \bibinfo {pages} {1655 } (\bibinfo {year} {2006})},\ \bibinfo {note} {2006
  Supplementary Non-Thematic Collection}\BibitemShut {NoStop}%
\bibitem [{\citenamefont {Klouda}\ and\ \citenamefont {Mikos}(2008)}]{Klouda}%
  \BibitemOpen
  \bibfield  {author} {\bibinfo {author} {\bibfnamefont {L.}~\bibnamefont
  {Klouda}}\ and\ \bibinfo {author} {\bibfnamefont {A.~G.}\ \bibnamefont
  {Mikos}},\ }\href@noop {} {\bibfield  {journal} {\bibinfo  {journal}
  {European Journal of Pharmaceutics and Biopharmaceutics}\ }\textbf {\bibinfo
  {volume} {68}},\ \bibinfo {pages} {34 } (\bibinfo {year} {2008})},\ \bibinfo
  {note} {interactive Polymers for Pharmaceutical and Biomedical
  Applications}\BibitemShut {NoStop}%
\bibitem [{\citenamefont {Purushotham}\ and\ \citenamefont
  {Ramanujan}(2010)}]{Purushotham}%
  \BibitemOpen
  \bibfield  {author} {\bibinfo {author} {\bibfnamefont {S.}~\bibnamefont
  {Purushotham}}\ and\ \bibinfo {author} {\bibfnamefont {R.}~\bibnamefont
  {Ramanujan}},\ }\href@noop {} {\bibfield  {journal} {\bibinfo  {journal}
  {Acta Biomaterialia}\ }\textbf {\bibinfo {volume} {6}},\ \bibinfo {pages}
  {502 } (\bibinfo {year} {2010})}\BibitemShut {NoStop}%
\bibitem [{\citenamefont {Wu}\ \emph {et~al.}(2011)\citenamefont {Wu},
  \citenamefont {Wang}, \citenamefont {Yu}, \citenamefont {Wang},\ and\
  \citenamefont {Chen}}]{Wu}%
  \BibitemOpen
  \bibfield  {author} {\bibinfo {author} {\bibfnamefont {Q.}~\bibnamefont
  {Wu}}, \bibinfo {author} {\bibfnamefont {L.}~\bibnamefont {Wang}}, \bibinfo
  {author} {\bibfnamefont {H.}~\bibnamefont {Yu}}, \bibinfo {author}
  {\bibfnamefont {J.}~\bibnamefont {Wang}}, \ and\ \bibinfo {author}
  {\bibfnamefont {Z.}~\bibnamefont {Chen}},\ }\href@noop {} {\bibfield
  {journal} {\bibinfo  {journal} {Chemical Reviews}\ }\textbf {\bibinfo
  {volume} {111}},\ \bibinfo {pages} {7855} (\bibinfo {year}
  {2011})}\BibitemShut {NoStop}%
\bibitem [{\citenamefont {Bernfeld}\ and\ \citenamefont
  {Wan}(1963)}]{Bernfeld}%
  \BibitemOpen
  \bibfield  {author} {\bibinfo {author} {\bibfnamefont {P.}~\bibnamefont
  {Bernfeld}}\ and\ \bibinfo {author} {\bibfnamefont {J.}~\bibnamefont {Wan}},\
  }\href@noop {} {\bibfield  {journal} {\bibinfo  {journal} {Science}\ }\textbf
  {\bibinfo {volume} {142}},\ \bibinfo {pages} {678} (\bibinfo {year}
  {1963})}\BibitemShut {NoStop}%
\bibitem [{\citenamefont {Horbett}\ \emph {et~al.}(1984)\citenamefont
  {Horbett}, \citenamefont {Kost},\ and\ \citenamefont {Ratner}}]{Horbett}%
  \BibitemOpen
  \bibfield  {author} {\bibinfo {author} {\bibfnamefont {T.~A.}\ \bibnamefont
  {Horbett}}, \bibinfo {author} {\bibfnamefont {J.}~\bibnamefont {Kost}}, \
  and\ \bibinfo {author} {\bibfnamefont {B.~D.}\ \bibnamefont {Ratner}},\
  }\enquote {\bibinfo {title} {Swelling behavior of glucose sensitive
  membranes},}\ in\ \href@noop {} {\emph {\bibinfo {booktitle} {Polymers as
  Biomaterials}}}\ (\bibinfo  {publisher} {Springer US},\ \bibinfo {address}
  {Boston, MA},\ \bibinfo {year} {1984})\ pp.\ \bibinfo {pages}
  {193--207}\BibitemShut {NoStop}%
\bibitem [{\citenamefont {Klumb}\ and\ \citenamefont {Horbett}(1992)}]{Klumb}%
  \BibitemOpen
  \bibfield  {author} {\bibinfo {author} {\bibfnamefont {L.~A.}\ \bibnamefont
  {Klumb}}\ and\ \bibinfo {author} {\bibfnamefont {T.~A.}\ \bibnamefont
  {Horbett}},\ }\href@noop {} {\bibfield  {journal} {\bibinfo  {journal}
  {Journal of Controlled Release}\ }\textbf {\bibinfo {volume} {18}},\ \bibinfo
  {pages} {59 } (\bibinfo {year} {1992})}\BibitemShut {NoStop}%
\bibitem [{\citenamefont {Hoffman}(2013)}]{Hoffman}%
  \BibitemOpen
  \bibfield  {author} {\bibinfo {author} {\bibfnamefont {A.~S.}\ \bibnamefont
  {Hoffman}},\ }\href@noop {} {\bibfield  {journal} {\bibinfo  {journal}
  {Advanced Drug Delivery Reviews}\ }\textbf {\bibinfo {volume} {65}},\
  \bibinfo {pages} {10 } (\bibinfo {year} {2013})},\ \bibinfo {note} {advanced
  Drug Delivery: Perspectives and Prospects}\BibitemShut {NoStop}%
\bibitem [{\citenamefont {Qiu}\ and\ \citenamefont {Park}(2001)}]{Qiu}%
  \BibitemOpen
  \bibfield  {author} {\bibinfo {author} {\bibfnamefont {Y.}~\bibnamefont
  {Qiu}}\ and\ \bibinfo {author} {\bibfnamefont {K.}~\bibnamefont {Park}},\
  }\href@noop {} {\bibfield  {journal} {\bibinfo  {journal} {Advanced Drug
  Delivery Reviews}\ }\textbf {\bibinfo {volume} {53}},\ \bibinfo {pages} {321
  } (\bibinfo {year} {2001})},\ \bibinfo {note} {triggering in Drug Delivery
  Systems}\BibitemShut {NoStop}%
\bibitem [{\citenamefont {Wang}\ \emph {et~al.}(2008)\citenamefont {Wang},
  \citenamefont {Wang}, \citenamefont {Detamore},\ and\ \citenamefont
  {Berkland}}]{Wang}%
  \BibitemOpen
  \bibfield  {author} {\bibinfo {author} {\bibfnamefont {Q.}~\bibnamefont
  {Wang}}, \bibinfo {author} {\bibfnamefont {L.}~\bibnamefont {Wang}}, \bibinfo
  {author} {\bibfnamefont {M.}~\bibnamefont {Detamore}}, \ and\ \bibinfo
  {author} {\bibfnamefont {C.}~\bibnamefont {Berkland}},\ }\href@noop {}
  {\bibfield  {journal} {\bibinfo  {journal} {Advanced Materials}\ }\textbf
  {\bibinfo {volume} {20}},\ \bibinfo {pages} {236} (\bibinfo {year}
  {2008})}\BibitemShut {NoStop}%
\bibitem [{\citenamefont {Bell}\ \emph {et~al.}(2006)\citenamefont {Bell},
  \citenamefont {Carrick}, \citenamefont {Katta}, \citenamefont {Jin},
  \citenamefont {Ingham}, \citenamefont {Aggeli}, \citenamefont {Boden},
  \citenamefont {Waigh},\ and\ \citenamefont {Fisher}}]{Bell}%
  \BibitemOpen
  \bibfield  {author} {\bibinfo {author} {\bibfnamefont {C.~J.}\ \bibnamefont
  {Bell}}, \bibinfo {author} {\bibfnamefont {L.~M.}\ \bibnamefont {Carrick}},
  \bibinfo {author} {\bibfnamefont {J.}~\bibnamefont {Katta}}, \bibinfo
  {author} {\bibfnamefont {Z.}~\bibnamefont {Jin}}, \bibinfo {author}
  {\bibfnamefont {E.}~\bibnamefont {Ingham}}, \bibinfo {author} {\bibfnamefont
  {A.}~\bibnamefont {Aggeli}}, \bibinfo {author} {\bibfnamefont
  {N.}~\bibnamefont {Boden}}, \bibinfo {author} {\bibfnamefont {T.~A.}\
  \bibnamefont {Waigh}}, \ and\ \bibinfo {author} {\bibfnamefont
  {J.}~\bibnamefont {Fisher}},\ }\href@noop {} {\bibfield  {journal} {\bibinfo
  {journal} {Journal of Biomedical Materials Research Part A}\ }\textbf
  {\bibinfo {volume} {78A}},\ \bibinfo {pages} {236} (\bibinfo {year}
  {2006})}\BibitemShut {NoStop}%
\bibitem [{\citenamefont {Guvendiren}\ \emph {et~al.}(2012)\citenamefont
  {Guvendiren}, \citenamefont {Lu},\ and\ \citenamefont
  {Burdick}}]{Guvendiren}%
  \BibitemOpen
  \bibfield  {author} {\bibinfo {author} {\bibfnamefont {M.}~\bibnamefont
  {Guvendiren}}, \bibinfo {author} {\bibfnamefont {H.~D.}\ \bibnamefont {Lu}},
  \ and\ \bibinfo {author} {\bibfnamefont {J.~A.}\ \bibnamefont {Burdick}},\
  }\href@noop {} {\bibfield  {journal} {\bibinfo  {journal} {Soft Matter}\
  }\textbf {\bibinfo {volume} {8}},\ \bibinfo {pages} {260} (\bibinfo {year}
  {2012})}\BibitemShut {NoStop}%
\bibitem [{\citenamefont {Gutowska}\ \emph {et~al.}(2001)\citenamefont
  {Gutowska}, \citenamefont {Jeong},\ and\ \citenamefont
  {Jasionowski}}]{Gutowska}%
  \BibitemOpen
  \bibfield  {author} {\bibinfo {author} {\bibfnamefont {A.}~\bibnamefont
  {Gutowska}}, \bibinfo {author} {\bibfnamefont {B.}~\bibnamefont {Jeong}}, \
  and\ \bibinfo {author} {\bibfnamefont {M.}~\bibnamefont {Jasionowski}},\
  }\href@noop {} {\bibfield  {journal} {\bibinfo  {journal} {The Anatomical
  Record}\ }\textbf {\bibinfo {volume} {263}},\ \bibinfo {pages} {342}
  (\bibinfo {year} {2001})}\BibitemShut {NoStop}%
\bibitem [{\citenamefont {Choat}\ \emph {et~al.}(2008)\citenamefont {Choat},
  \citenamefont {Cobb},\ and\ \citenamefont {Jansen}}]{Choat2008}%
  \BibitemOpen
  \bibfield  {author} {\bibinfo {author} {\bibfnamefont {B.}~\bibnamefont
  {Choat}}, \bibinfo {author} {\bibfnamefont {A.~R.}\ \bibnamefont {Cobb}}, \
  and\ \bibinfo {author} {\bibfnamefont {S.}~\bibnamefont {Jansen}},\
  }\href@noop {} {\bibfield  {journal} {\bibinfo  {journal} {New phytologist}\
  }\textbf {\bibinfo {volume} {177}},\ \bibinfo {pages} {608} (\bibinfo {year}
  {2008})}\BibitemShut {NoStop}%
\bibitem [{\citenamefont {Beebe}\ \emph {et~al.}(2000)\citenamefont {Beebe},
  \citenamefont {Moore}, \citenamefont {Bauer}, \citenamefont {Yu},
  \citenamefont {Liu}, \citenamefont {Devadoss},\ and\ \citenamefont
  {Jo}}]{Beebe}%
  \BibitemOpen
  \bibfield  {author} {\bibinfo {author} {\bibfnamefont {D.~J.}\ \bibnamefont
  {Beebe}}, \bibinfo {author} {\bibfnamefont {J.~S.}\ \bibnamefont {Moore}},
  \bibinfo {author} {\bibfnamefont {J.~M.}\ \bibnamefont {Bauer}}, \bibinfo
  {author} {\bibfnamefont {Q.}~\bibnamefont {Yu}}, \bibinfo {author}
  {\bibfnamefont {R.~H.}\ \bibnamefont {Liu}}, \bibinfo {author} {\bibfnamefont
  {C.}~\bibnamefont {Devadoss}}, \ and\ \bibinfo {author} {\bibfnamefont
  {B.-H.}\ \bibnamefont {Jo}},\ }\href@noop {} {\bibfield  {journal} {\bibinfo
  {journal} {Nature}\ }\textbf {\bibinfo {volume} {404}} (\bibinfo {year}
  {2000})}\BibitemShut {NoStop}%
\bibitem [{\citenamefont {Pedley}\ and\ \citenamefont
  {Fischbarg}(1978)}]{Pedley1978}%
  \BibitemOpen
  \bibfield  {author} {\bibinfo {author} {\bibfnamefont {T.}~\bibnamefont
  {Pedley}}\ and\ \bibinfo {author} {\bibfnamefont {J.}~\bibnamefont
  {Fischbarg}},\ }\href@noop {} {\bibfield  {journal} {\bibinfo  {journal}
  {Journal of theoretical biology}\ }\textbf {\bibinfo {volume} {70}},\
  \bibinfo {pages} {427} (\bibinfo {year} {1978})}\BibitemShut {NoStop}%
\bibitem [{\citenamefont {Carroll}\ \emph {et~al.}(2014)\citenamefont
  {Carroll}, \citenamefont {Jensen}, \citenamefont {Parsa}, \citenamefont
  {Holbrook},\ and\ \citenamefont {Weitz}}]{Carroll2014}%
  \BibitemOpen
  \bibfield  {author} {\bibinfo {author} {\bibfnamefont {N.~J.}\ \bibnamefont
  {Carroll}}, \bibinfo {author} {\bibfnamefont {K.~H.}\ \bibnamefont {Jensen}},
  \bibinfo {author} {\bibfnamefont {S.}~\bibnamefont {Parsa}}, \bibinfo
  {author} {\bibfnamefont {N.~M.}\ \bibnamefont {Holbrook}}, \ and\ \bibinfo
  {author} {\bibfnamefont {D.~A.}\ \bibnamefont {Weitz}},\ }\href@noop {}
  {\bibfield  {journal} {\bibinfo  {journal} {Langmuir}\ }\textbf {\bibinfo
  {volume} {30}},\ \bibinfo {pages} {4868} (\bibinfo {year}
  {2014})}\BibitemShut {NoStop}%
\bibitem [{\citenamefont {Bruus}(2007)}]{bruus}%
  \BibitemOpen
  \bibfield  {author} {\bibinfo {author} {\bibfnamefont {H.}~\bibnamefont
  {Bruus}},\ }\href@noop {} {\emph {\bibinfo {title} {Theoretical
  Microfluidics}}}\ (\bibinfo  {publisher} {Oxford University Press},\ \bibinfo
  {year} {2007})\BibitemShut {NoStop}%
\end{thebibliography}
%



\end{document}